# Multi-Parameter Analysis of Li-ion Battery Degradation: Integrating Optical Fiber Sensing with Differential State of Health Metrics

Idris Temitope Bello[a,b*], Hassan Raza[c], Madithedu Muneeswara[a], Neha Tewari[a], Yin Nee Cheung[a], Tobi Alabi Michael[a], Ridwan Taiwo[d], Fiske Lin[a]

[a]Centre for Advances in Reliability and Safety (CAiRS), Hong Kong, SAR, China

[b]School of Aerospace and Mechanical Engineering, University of Oklahoma, Norman, OK 73019 USA.

[c]Department of Mechanical Engineering, The Hong Kong Polytechnic University, SAR, China

[d]Department of Building and Real Estate, The Hong Kong Polytechnic University, SAR, China

[*]Corresponding author: itbello@ou.edu

# Abstract

The reliability and safety of Lithium-ion batteries (LiBs) are of great concern in the energy storage industry. Nevertheless, the real-time monitoring of their degradation remains challenging due to limited quantitative metrics available during cycling. This study addresses this limitation by employing a novel approach that combines external optical fiber sensing with advanced data analysis techniques to comprehensively assess battery health. We engineered a non-invasive optical sensing platform using tandem pairs of polymeric and silica-based fiber Bragg grating (FBG) sensors affixed to the external surface of commercial Li-ion button cells, enabling simultaneous, real-time monitoring of device-level volume changes and thermal events over 600 cycles. Our analysis incorporated differential techniques to estimate the battery's state of health (SOH) based on capacity, strain, and temperature variations with respect to voltage. Additionally, we implemented and compared three deep learning models - Long Short-Term Memory (LSTM), Gated Recurrent Unit (GRU), and Artificial Neural Network (ANN) - to predict battery SOH over cycles. We were able to capture both continuous and spontaneous degradation events and provide unique insights into battery behavior across its lifecycle through differential analysis and new SOH metrics demonstrating high correlation with conventional measures. This multi-parameter approach, combining advanced sensing techniques with innovative data analysis and deep learning methods, contributes significantly to battery diagnostics, potentially improving reliability assessment, enhancing safety standards, and accelerating the development of more sustainable energy storage solutions.

# Introduction

Energy storage devices play a pivotal role in the global transition towards sustainable and renewable energy systems [1]. Among these, electrochemical energy storage technologies, particularly lithium-ion batteries (LiBs), have emerged as the cornerstone of portable electronics, electric vehicles, and grid-scale energy storage [1,2]. Their high energy density, long cycle life, and relatively low self-discharge rates have made them indispensable in our increasingly electrified world [3,4].

As the demand for more efficient and reliable energy storage solutions grows, so does the need for advanced monitoring and diagnostic tools [5]. The performance and safety of LiBs are critically dependent on their operating conditions and degradation state [6]. Traditional methods of battery health assessment often rely on simple metrics such as capacity fade or internal resistance increase [7]. However, these methods may not capture the complex interplay of chemical, thermal, and mechanical processes occurring within the battery during operation and aging [8,9].

Recent advancements in sensing technologies have opened new avenues for more comprehensive battery health monitoring [10–13]. Mechanical sensing approaches such as dilatometry, digital image correlation, wafer curvature, and strain gauges have been employed to measure expansion, strain, and other mechanical changes in batteries [14–16]. However, these methods often require custom cells or fixtures, limiting their applicability in commercial settings [5]. In contrast, fiber optic sensors offer a versatile alternative, capable of measuring multiple parameters like temperature, pressure and strain simultaneously, without interfering with the battery's electrochemical processes [17–22]. These sensors provide real-time data crucial for assessing battery state of health and predicting potential failure modes [17,20,21].

Parallel to the development of advanced sensing techniques, there has been significant progress in data analysis methods for battery diagnostics [5,6,9]. Differential voltage analysis (DVA), incremental capacity analysis (ICA), and machine learning approaches have been employed to extract more nuanced information from battery cycling data [16,23–26]. For instance, Chen et al. [24] developed a simple-to-parameterize quantitative diagnostic approach using differential voltage and incremental capacity analyses that can differentiate between loss of lithium inventory and loss of active materials in lithium-ion batteries. Carlos et al.[27] also highlighted how incremental capacity and differential voltage techniques can identify and quantify degradation modes in LiBs to enable better state of health diagnosis and prognosis. These techniques all aim to identify subtle changes in battery behavior that may be precursors to performance degradation or safety issues [28].

Despite these advancements, there remains a significant gap in integrating multi-parameter sensing with sophisticated data analysis for a more holistic understanding of battery health. Most studies focus on either sensing or data analysis separately, missing the opportunity to leverage the synergies between these approaches. Furthermore, the long-term reliability and practicality of advanced sensing systems in real-world battery applications have yet to be thoroughly demonstrated.

To address these limitations, our study presents a novel approach that combines external optical fiber sensing with advanced data analysis techniques to comprehensively assess battery health. We devised a non-invasive optical sensing platform for simultaneous, real-time monitoring of device-level volume changes and thermal events in commercial Li-ion button cells. We then implemented differential analysis techniques to estimate the battery's state of health (SOH) based on capacity, strain, and temperature variations with respect to voltage. Finally, we compared the performance of three deep learning models (LSTM, GRU, and ANN) in predicting battery SOH over cycles. By integrating multi-parameter sensing with innovative differential analysis methods and state-of-the-art deep learning models, we aim to provide a more comprehensive and nuanced assessment of battery health throughout its lifecycle. This study not only advances our understanding of battery degradation mechanisms but also the insights gained from this research could lead to more efficient battery management systems, enhanced safety protocols, and the development of next-generation energy storage solutions.

# Experimental Methodology

To achieve our objectives, we designed a multi-faceted experimental approach. Our methodology encompasses the fabrication and implementation of a novel optical sensing system, long-term cycling of commercial Li-ion button cells, and the development of advanced data analysis techniques. The following sections detail our experimental procedures, data collection methods, and analytical approaches.

## Cell Specifications and Cycling Protocols

This study utilized 1258 button cells manufactured by GP Batteries Ltd. The cells featured $LiCoO_2$ (LCO) cathodes and anodes composed of SiOx/C blend (SC) with silicon content exceeding 10%. The rated capacity of these cells was 70 mAh. The proprietary electrolyte formulation consisted of $LiPF_6$ dissolved in carbonate-based solvents, typical for rechargeable LiBs.

Prior to testing, the cells underwent a manufacturer-specified SEI (solid-electrolyte interphase) formation process. Subsequently, cycling tests were conducted at 0.5C charge and discharge rates until the cells reached the end-of-life criterion, defined as 80% of the initial capacity. All electrochemical measurements were performed using a LANHE CT3002A battery testing system in a Vötsch Industrietechnik climate chamber maintained at 25°C.

## Optical Fiber Sensor Design and Implementation

Two types of optical fibers were employed for sensing: a commercial silica single-mode fiber (SMF) (Silitec G657.B germanium-doped) and a custom Zeonex-based polymer optical fiber (POF). This dual-fiber approach was chosen to leverage the complementary properties of each fiber type. Silica fibers offer high stability and low optical losses, making them ideal for long-term measurements. In contrast, polymer fibers have higher strain sensitivity and better mechanical compatibility with the battery surface, allowing for more accurate strain measurements [21]. Fiber Bragg gratings (FBGs) of 3 mm length were inscribed in both fiber types using a phase-mask method with a 248 nm excimer laser (Braggstar M, Coherent Inc.).

The distinct thermal and strain sensitivities of these fibers (silica: 9.62 pm/°C and 0.839 pm/µɛ; polymer: -22.09 pm/°C and 1.52 pm/µɛ) enable the decoupling of temperature and strain effects using a matrix inversion algorithm. This dual-sensor approach provides a more robust and accurate measurement system, capable of distinguishing between thermal and mechanical changes in the battery during cycling.

## Sensor Mounting and Data Acquisition

Custom 3D-printed cell holders were designed to securely position the cells while allowing for expansion, as shown in **Figure 1**. The FBG sensors were affixed to the anode side of each cell using a commercial twin epoxy adhesive (Araldite fast-setting). A thin layer of thermal paste (STARS612, Balance Stars) was applied between the cell surface and the fibers to enhance heat transfer.

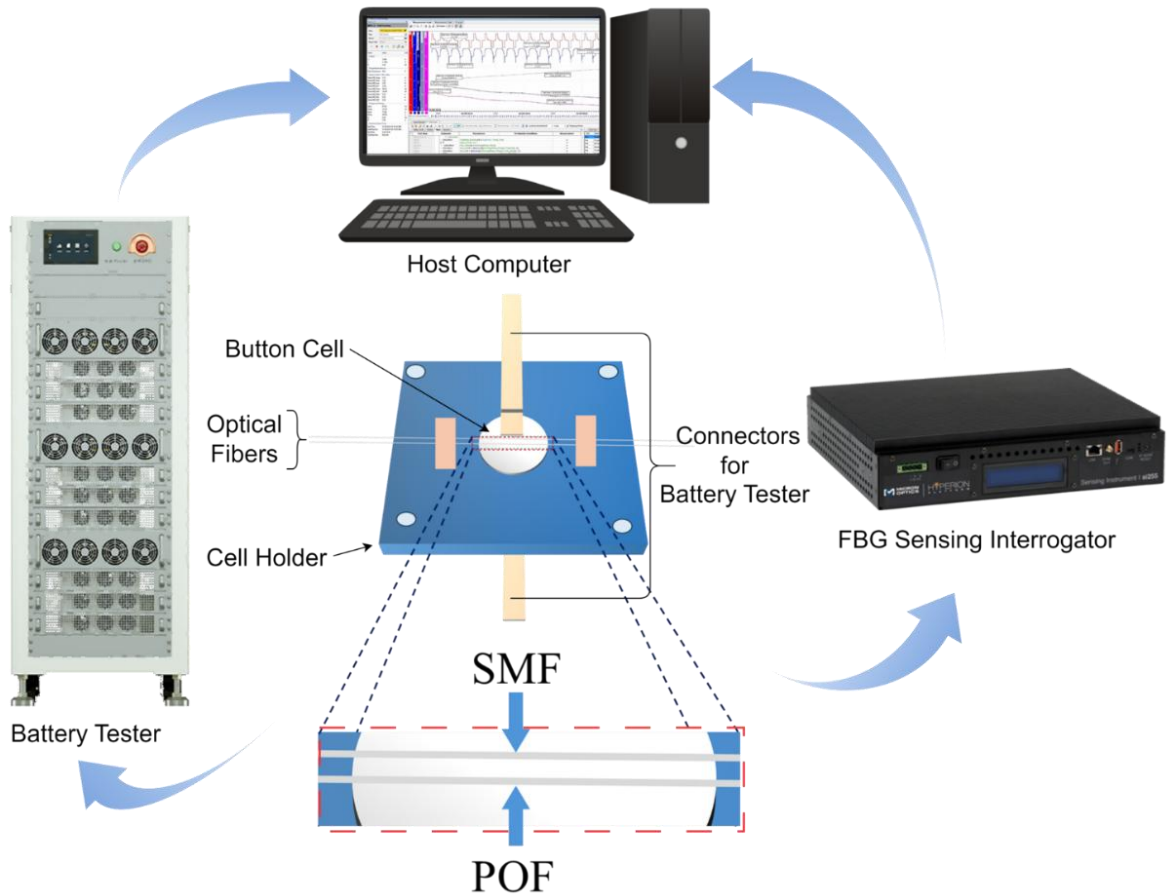

**Figure 1:** Schematic diagram of the multi-parameter battery monitoring system. The setup includes a button cell in a custom 3D-printed holder, optical fiber Bragg grating (FBG) sensors for strain and temperature measurements, an optical interrogator, and a battery tester, all integrated for comprehensive data acquisition and analysis.

FBG signals were recorded using a Luna Innovations Si155 optical interrogator. To ensure reliable electrical contact without impeding cell expansion, copper strips and conductive silver paste were used to connect the cell terminals to the battery tester.

## Calibration and Data Processing

Thermal calibration of the mounted sensors was performed between 25°C and 33°C to account for changes in thermal sensitivity due to the mounting process. The strain sensitivity remained constant after mounting. A custom Python script was developed to process the raw optical data and decouple the temperature and strain measurements.

## Data Analysis and Machine Learning

Differential analysis techniques were applied to the capacity, strain, and temperature data with respect to voltage. This analysis provides insights into the battery's SOH throughout its lifecycle by revealing subtle changes in battery behavior that may be precursors to performance degradation or safety issues. The differential parameters were calculated as follows:

$$(dQ/dV)_n = (\Delta Q/\Delta V)_n = (Q_{\{n+\Delta n\}} - Q_n)/(V_{\{n+\Delta n\}} - V_n) \qquad 1$$

$$(d\epsilon/dV)_n = (\Delta \epsilon/\Delta V)_n = (\epsilon_{\{n+\Delta n\}} - \epsilon_n)/(V_{\{n+\Delta n\}} - V_n) \qquad 2$$

$$(dT/dV)_n = (\Delta T/\Delta V)_n = (T_{\{n+\Delta n\}} - T_n)/(V_{\{n+\Delta n\}} - V_n) \qquad 3$$

In these equations, Q, $\epsilon$, T, and V represent capacity, strain, temperature, and voltage, respectively. The differentials (dQ/dV, dε/dV, and dT/dV) were calculated over a specified step size of Δn = 10, ensuring sufficient sampling points for analysis. The Locally Weighted Scatterplot Smoothing (LOWESS) method was then applied to the differential analysis data to provide clearer insights into the relationships between temperature, capacity, strain, and voltage during the battery cycles [29]. The LOWESS smoothing was performed for the derivatives dQ/dV, dε/dV, and dT/dV during charging and discharging of the cell, as shown in **Figure 2.**

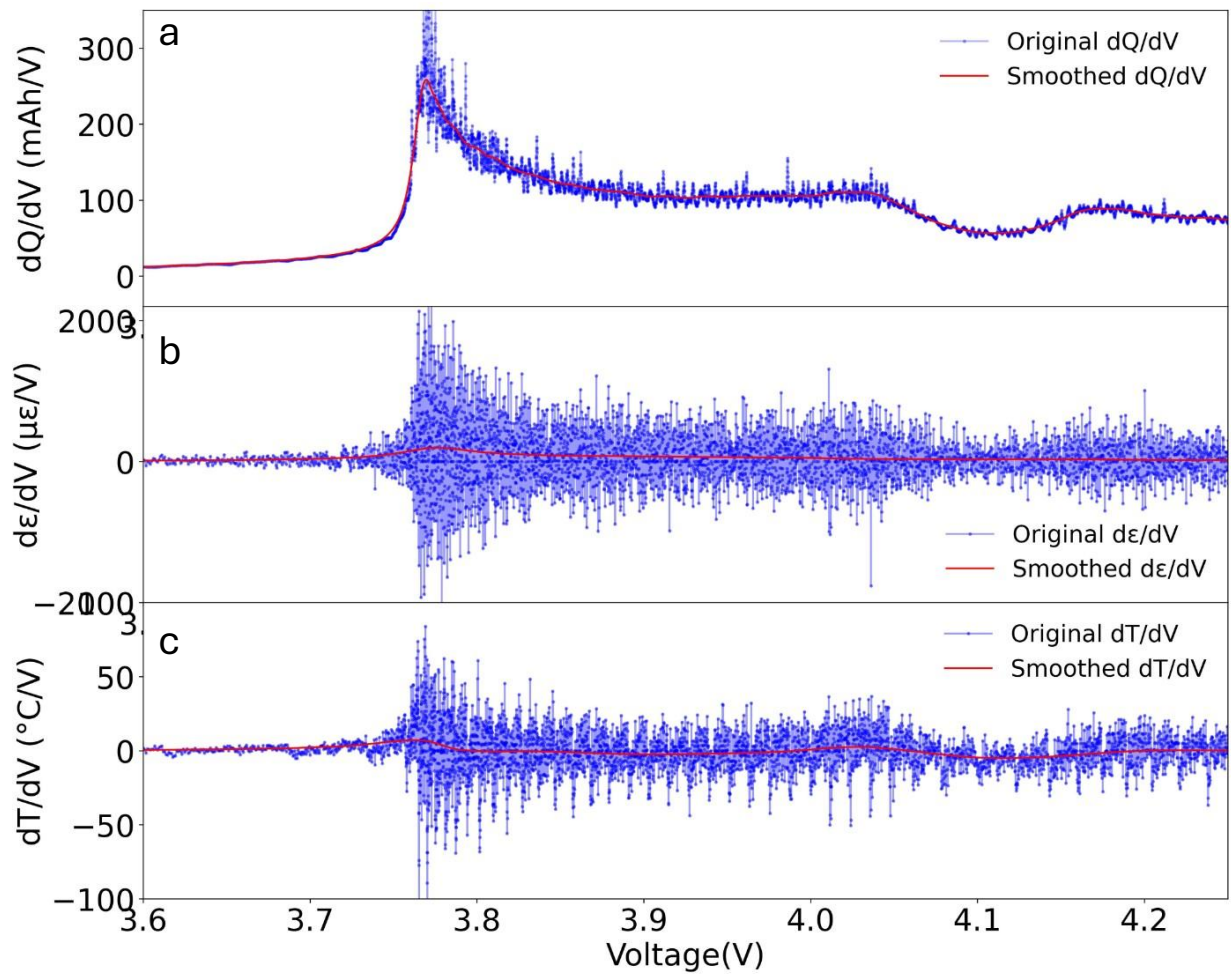

**Figure 2:** LOWESS smoothing applied to the various differentials. (a) Capacity (b) Strain and (c) Temperature.

This method is a non-parametric regression technique that fits multiple regressions in localized subsets of the data to produce a smooth curve through the points [30]. The general form of the LOWESS smoothing function can be expressed as:

$$\hat{y}i = \sum j = 1^n w_{ij} y_j \qquad 4$$

where $\hat{y}_i$ is the smoothed value at point (i), $y_j$ are the observed values, $w_{ij}$ are the weights assigned to each observed value $y_j$, which decrease with the distance from point (i). The weights $w_{ij}$ are calculated using a tricube weight function:

$$w_{ij} = \left(1 - \left(\frac{|x_i - x_j|}{D}\right)^3\right)^3 \qquad 5$$

where $x_i$ and $x_j$ are the predictor variable values (in this case, voltage), and (D) is the bandwidth parameter that controls the degree of smoothing.

This differential and smoothing approach allows us to identify and quantify specific electrochemical and physical processes occurring within the battery, such as phase transitions, structural changes, and thermal events. By tracking how these differential parameters evolve over the battery's lifetime, we can gain deeper insights into degradation mechanisms and potential indicators of battery health.

Additionally, three deep learning models - Long Short-Term Memory (LSTM), Gated Recurrent Unit (GRU), and Artificial Neural Network (ANN) - were implemented to predict the battery's capacity-based SOH over cycles. The models were trained on the multi-parameter data collected from the optical sensors and electrochemical measurements. Our experimental setup (**Figure 1**) and analysis methodology enabled a comprehensive, real-time assessment of battery health, combining high-precision optical sensing with advanced data analytics and machine learning techniques.

## Results and Discussion

Examining the complex interplay between electrochemical, mechanical, and thermal processes during battery operation and aging is crucial for developing more durable and highly performing energy storage solutions. To achieve this, we conducted a multi-parameter analysis of our SC-based button cell performance over an extended cycling period, as presented in **Figure 3**. A novel approach was employed, integrating external optical fiber sensors with advanced data analysis techniques to simultaneously monitor voltage, strain, and temperature profiles. **Figure 3a** showcases the long-term evolution of these parameters over 800 cycles, providing a holistic view of the cell degradation. The voltage profiles demonstrate a gradual narrowing of the operating window, indicative of increased internal resistance and loss of active material in the composite electrodes. The strain profiles reveal a progressive decline in amplitude, suggesting mechanical degradation of the electrode structure, likely due to repeated volume changes of the SC anode [31]. The temperature profiles, while exhibiting subtler changes, show slightly increased fluctuations over time, implying alterations in the cell's thermal behavior.

A detailed single-cycle analysis for cycle 5 is provided in **Figure 3b**, offering deeper insights into the intricate relationships between these parameters during charge and discharge processes. The evolution of voltage, strain, and temperature profiles with respect to state of charge (SOC) for representative cycles (**Figure 3c-e**) further emphasizes the progressive nature of the degradation.

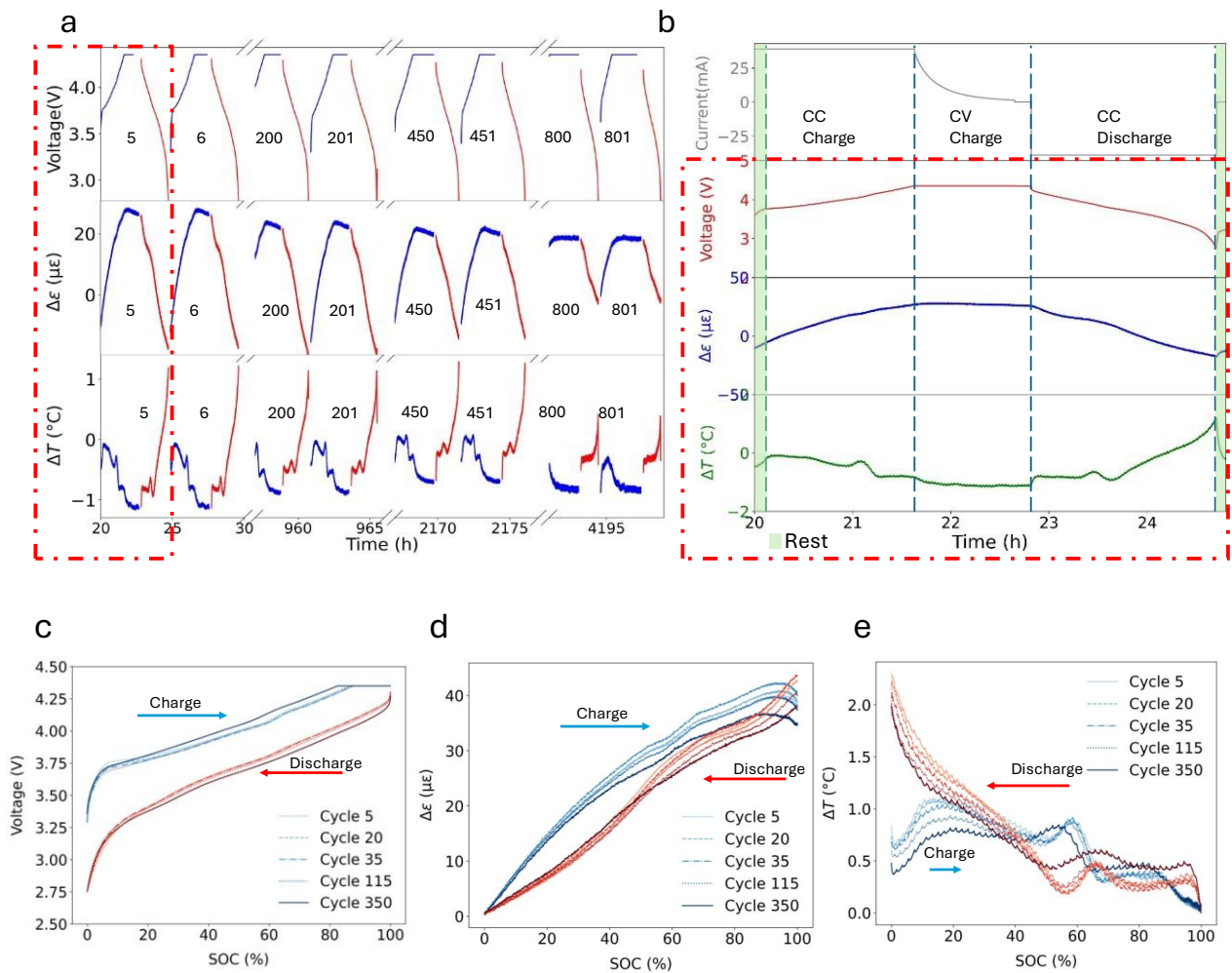

**Figure 3:** Multi-cycle characterization of SC-based button cell performance at 0.5C/0.5C. (a) Long-term voltage, strain, and temperature profiles demonstrating progressive cell degradation. (b) Detailed multi-parameter analysis during charge and discharge for cycle 5, highlighting the interplay between constant current (CC) and constant voltage (CV) phases. (c-e) Evolution of voltage, strain, and temperature profiles as a function of SOC for representative cycles (5, 20, 35, 115, and 350), demonstrating changes in electrochemical, mechanical, and thermal behavior over the cell's lifespan.

The observed decrease in strain amplitudes, coupled with capacity fade, underscores the significant role of mechanical degradation, likely driven by continuous volume changes in the SC anode, in the overall cell degradation. The evolving voltage profiles not only reflect a loss of active material but also suggest increasing internal resistance, potentially due to mechanical degradation and solid electrolyte interphase (SEI) growth on the SC anode [32]. The subtle changes in temperature profiles indicate evolving thermal behavior, likely influenced by changes in internal resistance and electrochemical processes as the cell ages.

To further understand the degradation mechanisms in the battery, particularly how mechanical and thermal changes correlate with capacity fade over time, we analyzed the maximum capacity, strain, and temperature per cycle over the cell's lifetime during charge and discharge (see **Figure 4a,b**) and at specific depth of discharge (DoD) (see **Figure 4c**). Capacity ($Q_{max}$) shows a gradual decline over cycles, indicating progressive capacity fade. Maximum strain follows a similar trend, experiencing a rapid progressive decrease in the early cycles followed by occasional spikes around 400 and 550 cycles before stabilizing. This behavior suggests that mechanical degradation may be more pronounced in early cycles. However, the unusual spikes or anomalies observed around 400 and 450 cycles may be attributed to sudden external disturbance or a measurement artifact. Maximum temperature change remains relatively constant with some fluctuations (around the same cycles in the strain profiles), indicating consistent thermal management despite capacity fading. The analysis at 50% DoD reveals a slight increasing trend in temperature change, a gradual decrease in strain change, and a steady decrease in capacity (**Figure 4c**). These trends suggest that the loss of active material or structural changes in the LCO and SC electrodes may reduce their ability to expand and contract during cycling. Specifically, the LCO cathode may experience a reduction in $Li^+$ intercalation capacity due to phase transitions and the formation of inactive phases [33,34]. Similarly, the SC anode may undergo volumetric changes leading to particle pulverization and loss of electrical contact [35]. These structural degradations contribute to increased internal resistance, which in turn leads to higher heat generation within the cell [36].

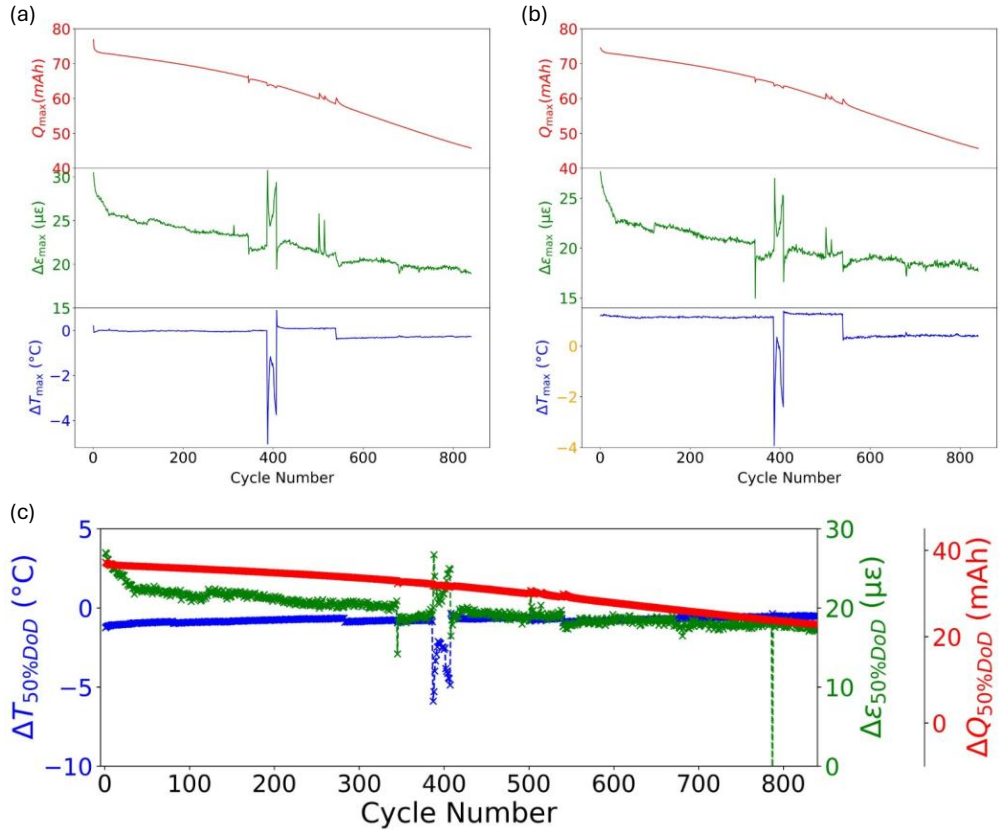

**Figure 4:** Temporal evolution of key battery parameters throughout the cell's lifecycle. (a) Maximum capacity, strain, and temperature changes during charging versus cycle number. (b) Corresponding parameter changes during discharging versus cycle number. (c) Capacity, strain, and temperature variations at 50% Depth of Discharge (DoD) across battery cycles, revealing distinct degradation patterns.

One of the key methods for assessing aging and diagnosing battery health is differential analysis [23,24]. Traditionally, differential voltage (dV/dQ) or capacity (dQ/dV) analysis are the most popular approaches, as they provide detailed insights into the electrochemical processes occurring within the battery during charge and discharge cycles [27]. However, these methods are limited to electrochemical parameters, which may not capture the full picture. To gain deeper insights into the various processes occurring in the battery, we performed a differential analysis of capacity, strain, and temperature with respect to voltage throughout the battery's lifetime, as shown in **Figure**

**5**. Given the extensive dataset of over 800 cycles, we focused on the dominant peak with the highest intensity to ensure a robust and computationally efficient analysis, which is crucial for practical industrial applications.

The dQ/dV plot reveals a prominent peak around 3.8V, indicating a significant phase transition or lithium insertion event in the LCO electrode (**Figure 5a)** [37,38]. This dominant peak was chosen for detailed analysis due to its high signal-to-noise ratio and strong correlation with the primary electrochemical process, making it an ideal indicator of overall battery health. The peak subsequently decreases in magnitude and broadens as the cycle number increases, suggesting a loss of crystallographic order or an increase in the amorphous nature of the LCO over time [36,39]. The coinciding spike in dε/dV (**Figure 5b)** suggests that this phase transition is accompanied by substantial structural changes in the LCO material. Subtle changes in dT/dV (**Figure 5c)**, particularly around the 3.8V region, indicate that these processes may be slightly exothermic. The strain and temperature responses become less pronounced in later cycles, indicating a general decline in the battery's ability to store and release energy efficiently. Increasing hysteresis between charge and discharge cycles points to growing irreversible losses, possibly due to side reactions, SEI (solid-electrolyte interphase) growth, or mechanical degradation [40,41].

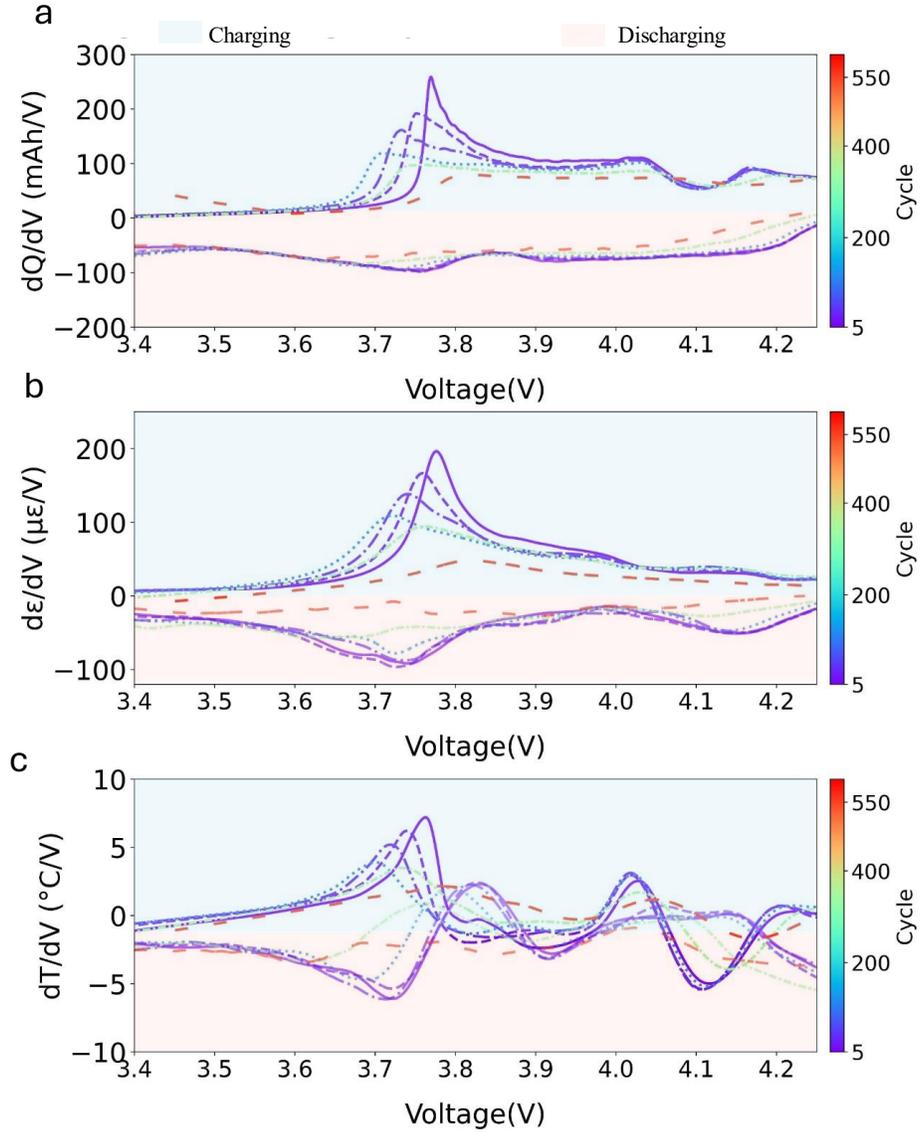

**Figure 5:** Comprehensive differential analysis of battery behavior across multiple charge-discharge cycles. Smoothed derivatives of (a) capacity (dQ/dV), (b) strain (dε/dV), and (c) temperature (dT/dV) with respect to voltage for cycles 5, 20, 35, 115, 350, 550, and 600, illustrating the evolution of battery characteristics over its lifetime.

To understand how the magnitudes of key electrochemical, mechanical, and thermal processes change over time, and identify potential indicators of battery aging and degradation, we analyzed the long-term trends in the peak intensities of the differential parameters (**Figure 5**) throughout the observed battery's lifecycle, as shown in **Figure 6** [42]. Focusing on peak intensities rather than positions provides a more sensitive and reliable measure of degradation, as it directly correlates with the capacity and kinetics of the primary electrochemical processes. During charge cycles, there is a rapid decrease in dQ/dV intensity in the first 100 cycles, followed by a more gradual decline. The dε/dV and dT/dV intensities showed similar initial drops but with more fluctuations. Discharge cycles exhibited more gradual changes compared to charge cycles. These trends suggest significant changes in the battery's behavior during the initial stages of its life, possibly related to SEI formation and initial structural changes, followed by a relatively stable state of operation for a significant portion of its life [43].

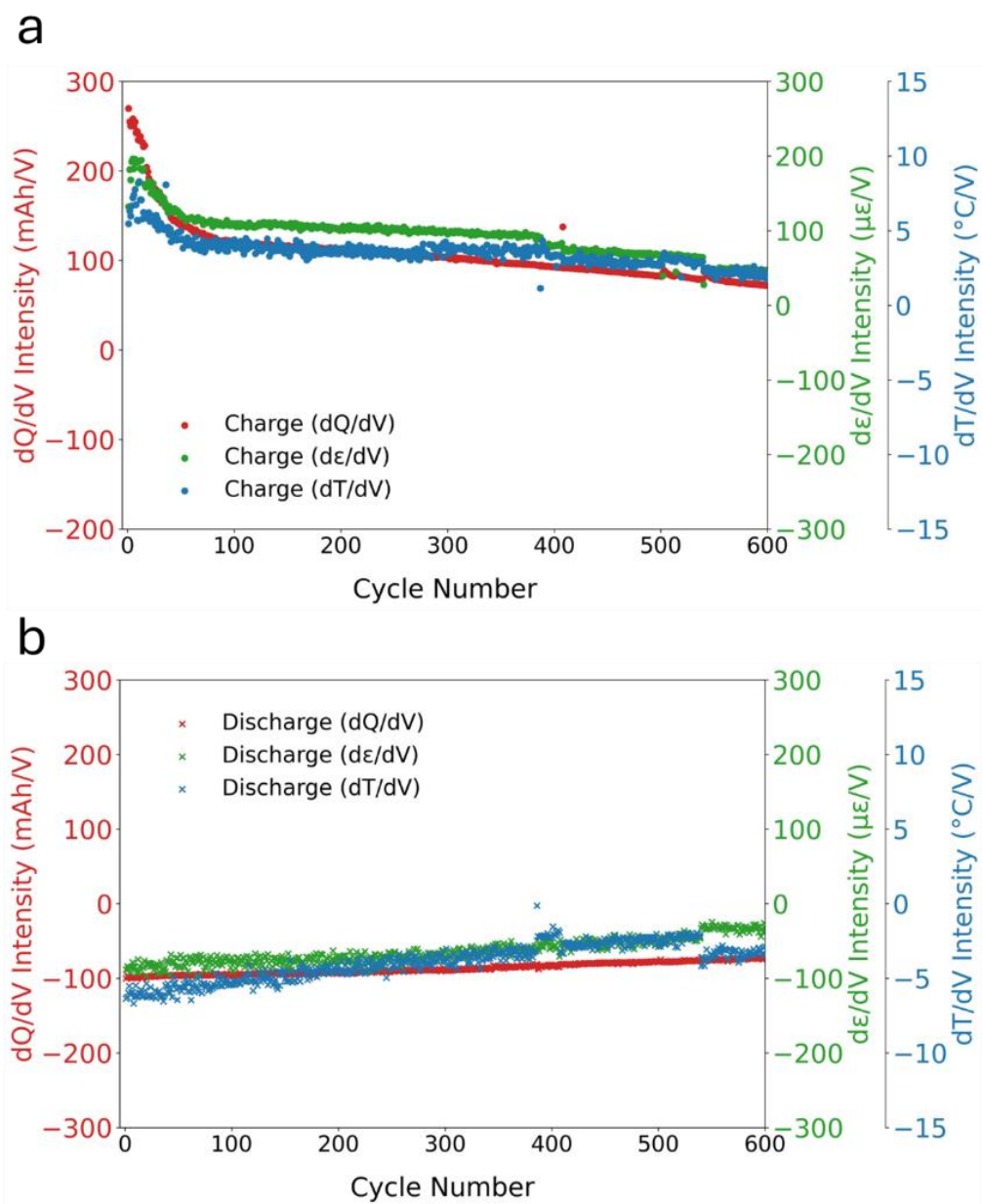

**Figure 6:** Long-term trends in differential intensity variations during (a) charge and (b) discharge cycles. Peak intensities of dQ/dV, dε/dV, and dT/dV plotted against cycle number, revealing distinct patterns of degradation in electrochemical, mechanical, and thermal processes.

To further understand how the voltages at which key electrochemical, mechanical, and thermal processes occur change over time, and identify potential shifts in reaction mechanisms or material properties, we tracked the corresponding voltage positions of the peak differential intensities from **Figure 6**, as shown in **Figure 7** [24]. While peak positions are generally less sensitive to degradation than intensities, they provide complementary information about thermodynamic changes in the system. During charge cycles, all parameters showed similar trends: an initial decrease, followed by a minimum around cycle 100, then a gradual increase. Discharge cycles exhibited a general downward trend over time with much more scatter in the data. These findings indicate that the voltage ranges for key electrochemical, mechanical, and thermal processes are intricately linked and evolve together over the battery's lifetime. The downward trend in discharge peak voltages suggests that key processes are occurring at progressively lower voltages, possibly due to increased polarization or changes in the properties of the LCO and SC. The consistent ordering of peak voltages ($d\varepsilon/dV > dQ/dV > dT/dV$) during charging suggests a specific sequence of events: mechanical changes in the SC anode precede or drive capacity changes in the LCO cathode, which then lead to thermal effects.

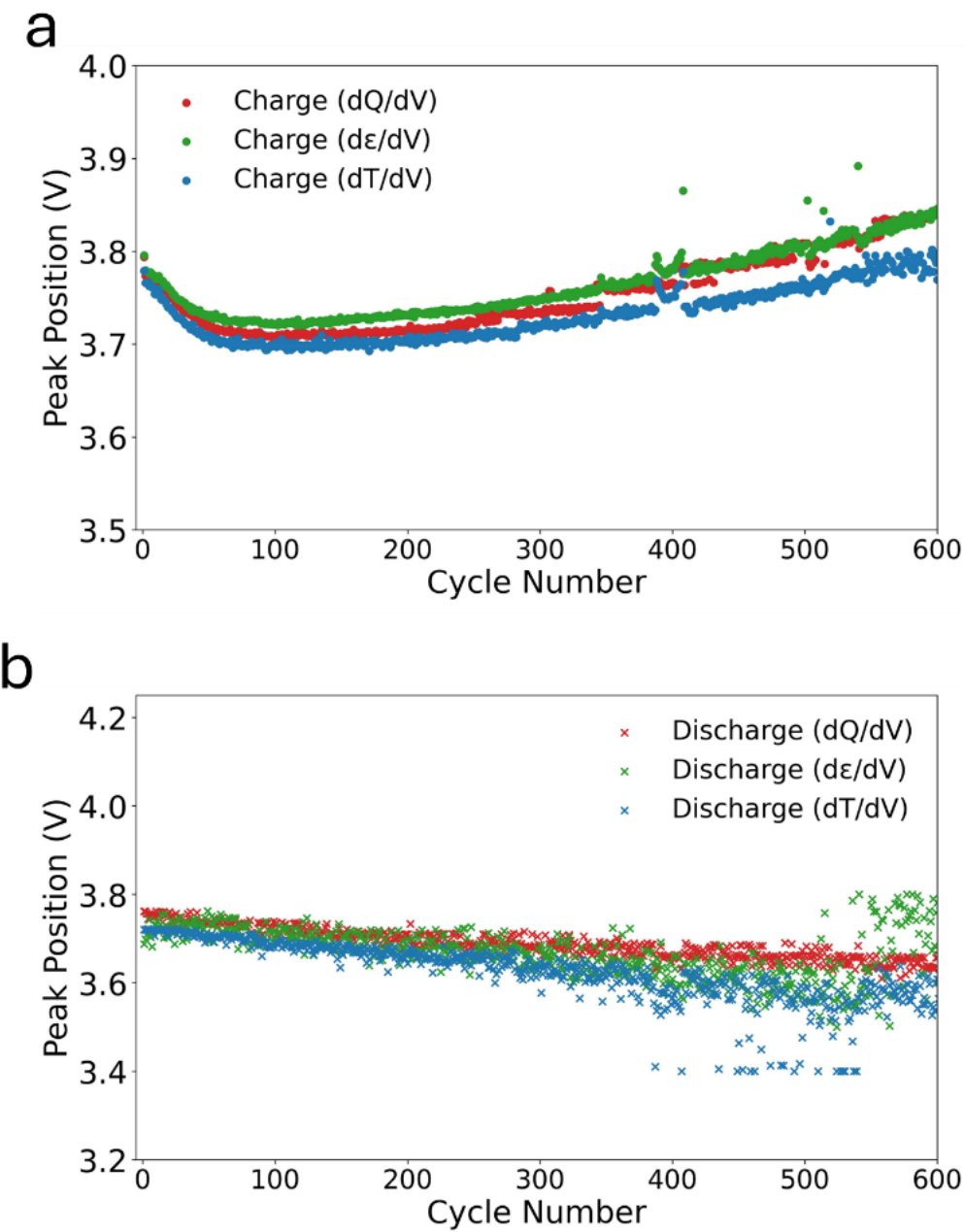

**Figure 7:** Evolution of differential intensity peak positions throughout the battery's lifecycle. Voltage positions corresponding to maximum dQ/dV, dε/dV, and dT/dV during (a) charge and (b) discharge cycles, providing insights into shifts in key electrochemical and physical processes over time.

The downward trend in discharge peak voltages indicates that key processes are occurring at progressively lower voltages, due to increased polarization or changes in the properties of the LCO and SC electrodes [26]. Likewise, the greater scatter in discharge data suggests more variability in the discharge process, which could be related to inhomogeneous degradation or the influence of rest periods between cycles. By focusing on the dominant peak across all three differential parameters (dQ/dV, dε/dV, dT/dV), we ensure a consistent and comparable analysis throughout the battery's lifetime, providing a robust method for tracking degradation in practical, industrial applications.

In the day-to-day operation of batteries, reliability is crucial for ensuring customer satisfaction and safety. One key metric used to estimate battery reliability is the SOH metric [21,25,27,28]. To understand how different metrics (capacity, strain, and temperature) correlate with battery degradation, we assessed the potential of differential intensity-based SOH metrics (**Figure 8a,b**) as alternatives or complements to traditional capacity-based SOH (**Figure 8c**). The differential intensity-based SOH metrics were calculated using **equation 6.**

$$SOH_{dX/dV} = (I_{n,dX/dV} / I_{0,dX/dV}) \times 100 \qquad 6$$

Where X can be Q, ε, or T, and $I_n$ is the peak intensity of dX/dV at cycle n and $I_0$ is the peak intensity of dX/dV at the initial cycle.

A comparative analysis of different SOH metrics is presented in **Figure 8(d-f)**. We found a strong linear correlation between traditional capacity-based SOH ($SOH_Q$) and differential capacity-based $SOH_{dQ/dV}$ (**Figure 8d**), suggesting that the differential capacity method is a reliable alternative for estimating battery health. The more scattered correlations between strain-based $SOH_{dε/dV}$ or temperature-based $SOH_{dT/dV}$ and $SOH_{dQ/dV}$ indicate that these parameters might be capturing additional or distinct aspects of battery degradation compared to capacity-based metrics alone. This multi-parameter approach to SOH estimation could lead to more accurate predictions of battery lifetime and performance.

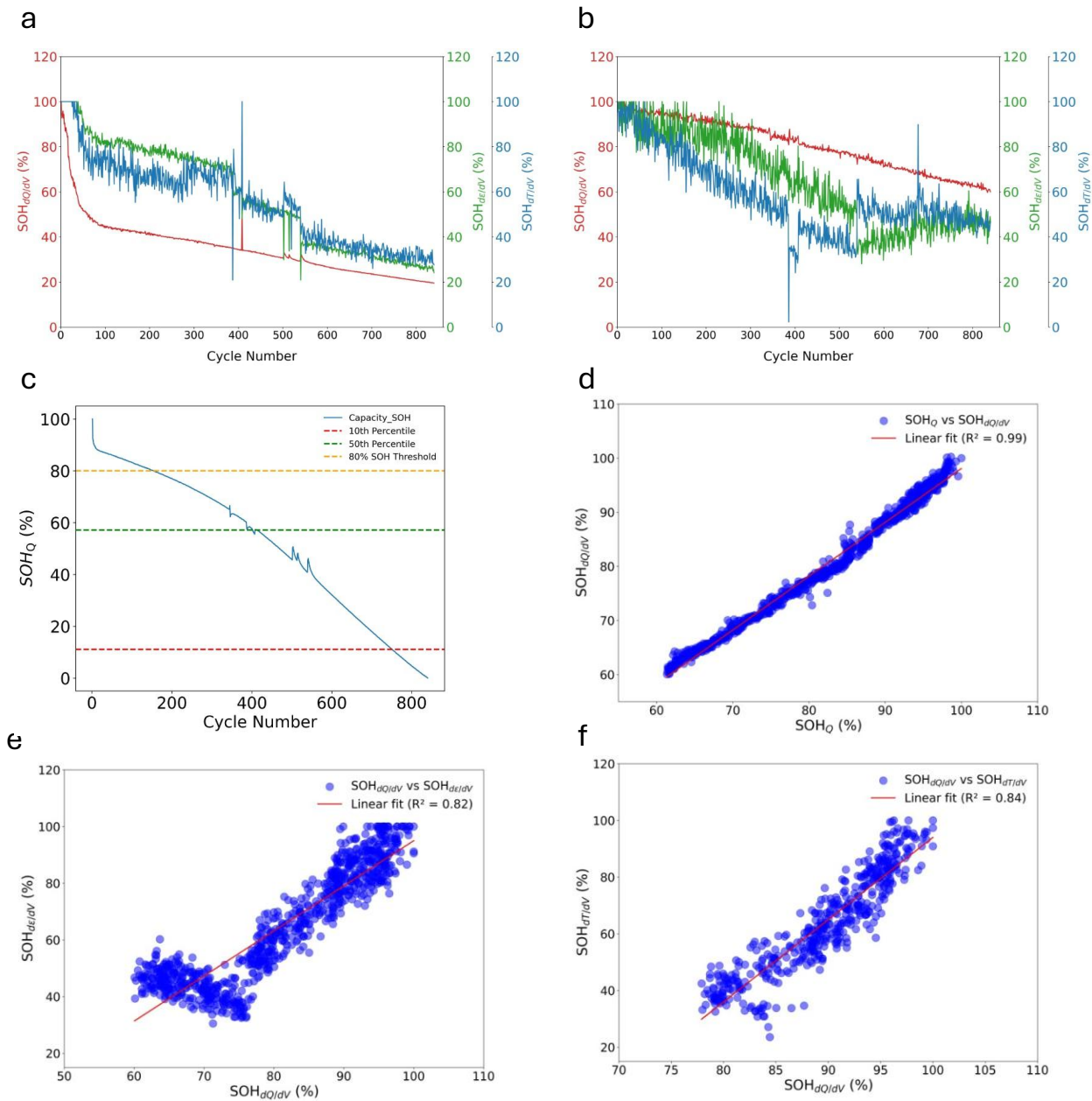

**Figure 8:** Comparative analysis of novel and traditional State SOH metrics. (a) SOH based on differential intensity analysis during charge cycles. (b) Corresponding SOH metrics during discharge cycles. (c) the traditional SOH (SOH$_Q$). Linear correlation between (c) the traditional SOH (SOH$_Q$) and differential capacity-based SOH (SOH$_{dQ/dV}$), (d) strain-based SOH SOH$_{d\varepsilon/dV}$ and SOH$_{dQ/dV}$, and (e) temperature-based (SOH) SOH$_{dT/dV}$ and SOH$_{dQ/dV}$.

Predicting or forecasting failures in energy systems is crucial for ensuring reliability and safety in both industrial and everyday operations. Deep learning models have shown promise for time series analysis, but achieving optimal results requires identifying potential redundancies or synergies among different parameters [44–47]. To address this, we examined the interrelationships between various battery parameters and identified key features that significantly influence battery health and performance. **Figure 9** presents a color-coded Pearson correlation matrix for 18 key battery parameters: cycle number, current, voltage, capacity, temperature change, strain change, SOH, and row count. We calculated Pearson correlation coefficients between all pairs of parameters using the entire dataset collected throughout the battery's lifetime. The resulting correlation matrix was visualized as a heatmap, with correlation coefficients ranging from -1 to 1, where red indicates positive correlations and blue indicates negative correlations.

The correlation matrix (**Figure 9**) reveals several important relationships between battery parameters. The strong negative correlation between Cycle Index and $SOH_Q$ (-0.98) confirms the well-established relationship between cycling and battery degradation [31,41]. This linear relationship suggests that simple cycle counting could be a reliable predictor of battery health for this cell chemistry and cycling conditions. Voltage parameters ($V_{mean}$, $V_{max}$) exhibited strong positive correlations with capacity and SOH (correlation coefficients > 0.9), indicating that voltage measurements could be reliable indicators of battery health. This finding suggests a potentially simpler method for SOH estimation in practical applications, where continuous capacity measurements may not be feasible.

Interestingly, we found a strong correlation between voltage and strain change ($\Delta\varepsilon_{max}$) (correlation coefficient = 0.85), suggesting that monitoring strain could provide valuable insights into the battery's voltage state and potentially its overall health. This correlation supports our multi-parameter approach to battery health monitoring and highlights the potential of mechanical measurements as early indicators of battery degradation.

Temperature change parameters ($\Delta T_{mean}$, $\Delta T_{max}$) demonstrated moderate positive correlations with each other but weaker correlations with other parameters. This suggests that thermal effects may have a more complex, non-linear relationship with battery degradation, warranting further investigation into the role of thermal dynamics in battery health.

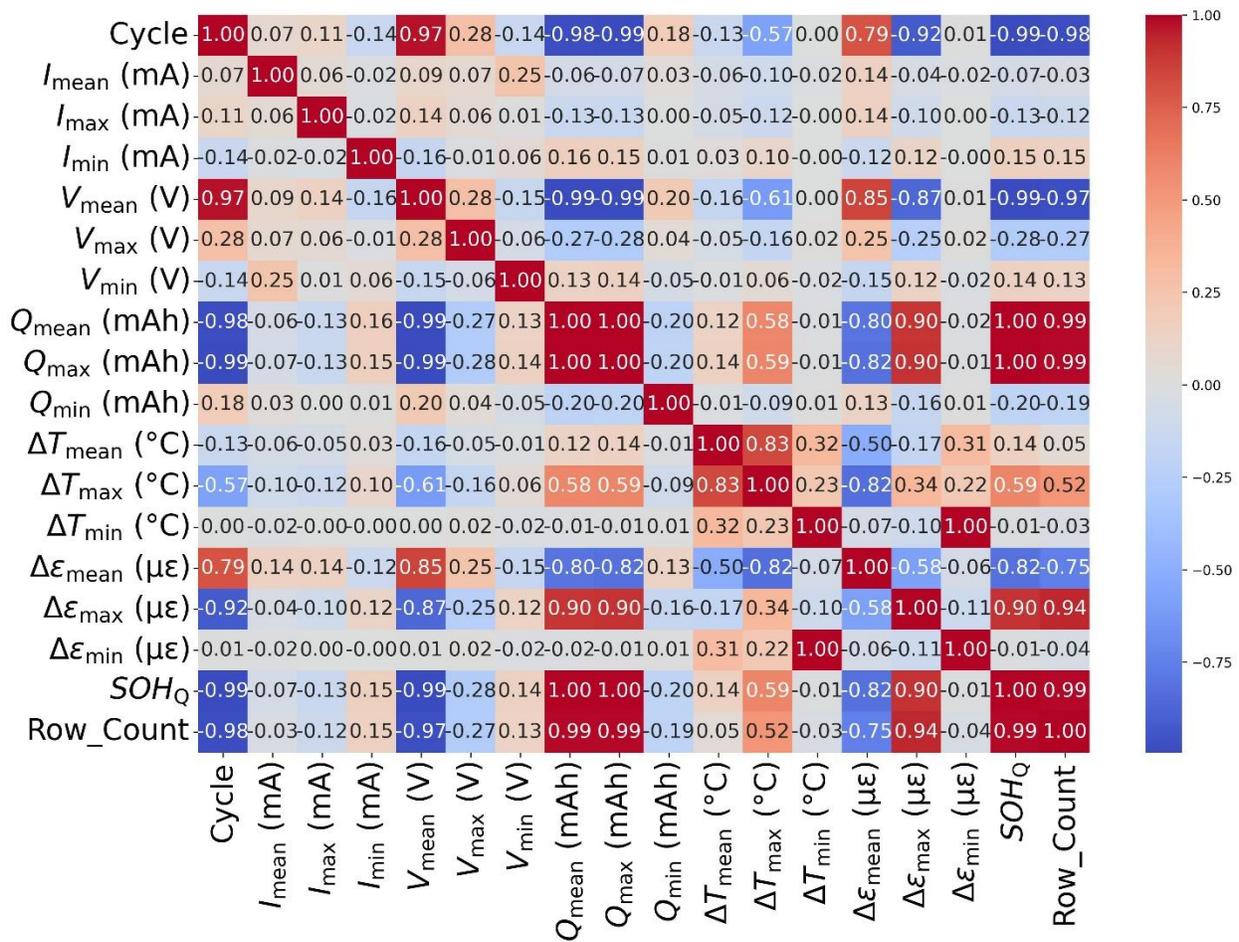

**Figure 9:** Pearson correlation matrix of key battery parameters and features used in deep learning models. The heatmap illustrates the interdependencies between Cycle Index, Current, Voltage, Capacity (Q), Temperature Change ($\Delta T$), Strain Change ($\Delta \varepsilon$), and capacity-based SOH ($SOH_Q$).

These findings build upon our earlier differential analysis results by quantifying the relationships between various battery parameters across the entire dataset. The strong correlations observed between mechanical (strain), electrochemical (capacity, voltage), and health (SOH) parameters reinforce the value of our multi-parameter sensing approach. Furthermore, these correlations provide valuable guidance for feature selection in our deep-learning models, potentially improving their predictive accuracy.

**Figure 10** presents the various deep-learning models we employed in predicting the SOH of the battery. We implemented three different neural network architectures: Gated Recurrent Unit (GRU), Artificial Neural Network (ANN), and Long Short-Term Memory (LSTM). From **Figure 10 (a-c),** all the models indicate rapid initial convergence, with loss decreasing sharply in the first few epochs. GRU and LSTM converge slightly slower than ANN but achieve stable final loss values. The three examined models demonstrated excellent prediction accuracy, closely tracking the actual SOH curve and capturing the non-linear degradation pattern of the battery SOH. This indicates that the models have learned generalizable patterns of battery degradation, applicable to both the early and late stages of the battery lifecycle. The success of GRU and LSTM models suggests that there are significant temporal dependencies in battery degradation processes, which these recurrent architectures can effectively learn and utilize.

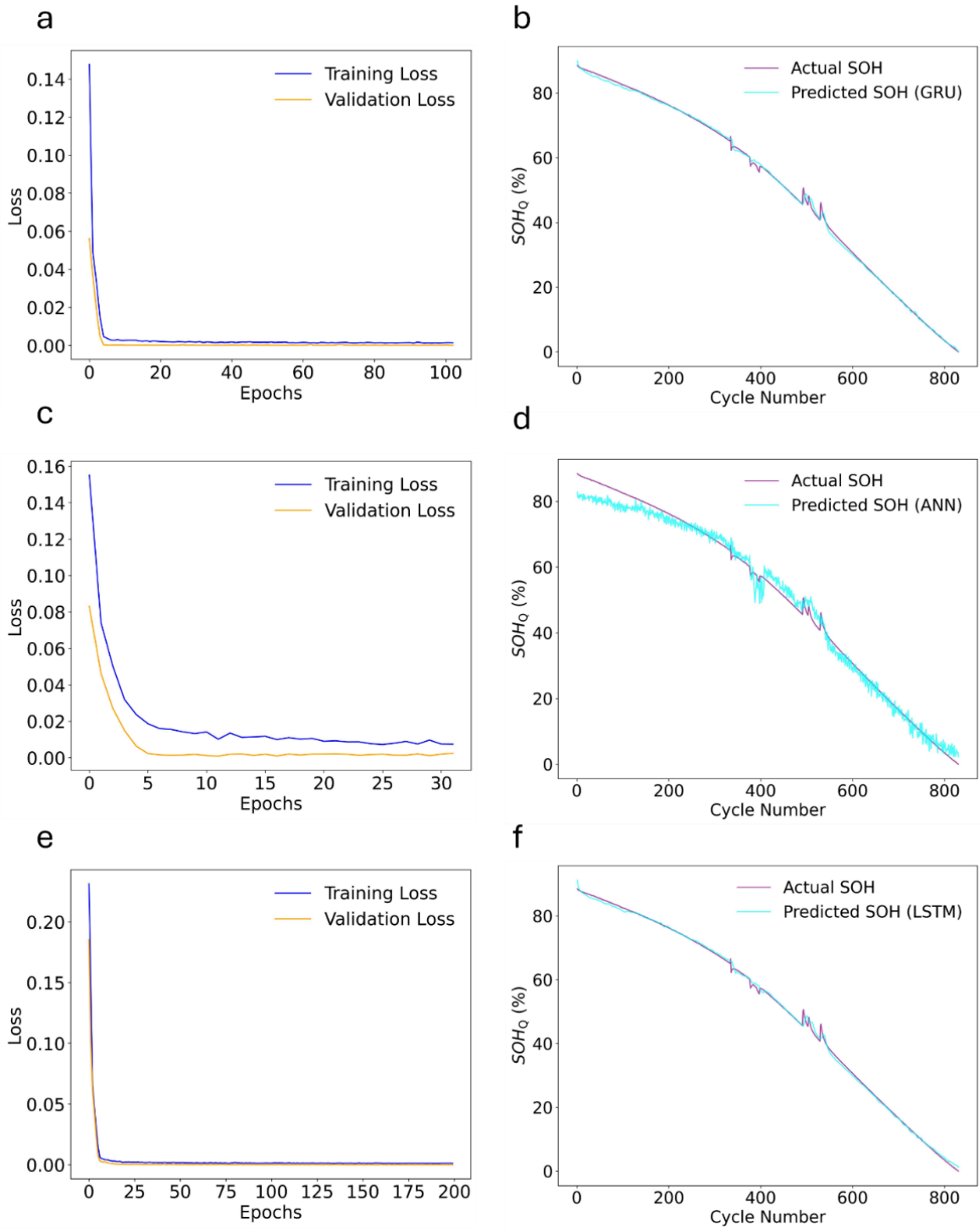

**Figure 10:** Performance evaluation of deep learning models for battery SOH prediction. (a-c) Training and validation loss curves over epochs for GRU, ANN, and LSTM models, respectively. (d-f) Actual vs. predicted SOH over battery cycles for GRU, ANN, and LSTM models, demonstrating their predictive capabilities.

We finally quantitatively compared the performance of these deep-learning models using three key metrics: Mean Squared Error (MSE), Root Mean Squared Error (RMSE), and Mean Absolute Error (MAE), (see **Equations 9-11)**, as shown in **Figure 11**. The results provide a comprehensive view of model performance for our battery SOH prediction.

The GRU model demonstrates superior performance across all metrics (MSE: 0.64, RMSE: 0.80, MAE: 0.58), highlighting its effectiveness in capturing and utilizing patterns in battery degradation data. This suggests that GRU's simpler architecture strikes an optimal balance between complexity and performance for this task. The LSTM model, while strong, shows slightly higher error rates (MSE: 1.05, RMSE: 1.03, MAE: 0.77), challenging the assumption that its additional complexity always provides an advantage in temporal sequence modeling. In contrast, the ANN model exhibits significantly higher error rates (MSE: 4.87, RMSE: 2.21, MAE: 1.81), indicating that simple feedforward architectures struggle with the complex, time-dependent nature of battery degradation, underscoring the importance of considering temporal dynamics in battery health modeling.

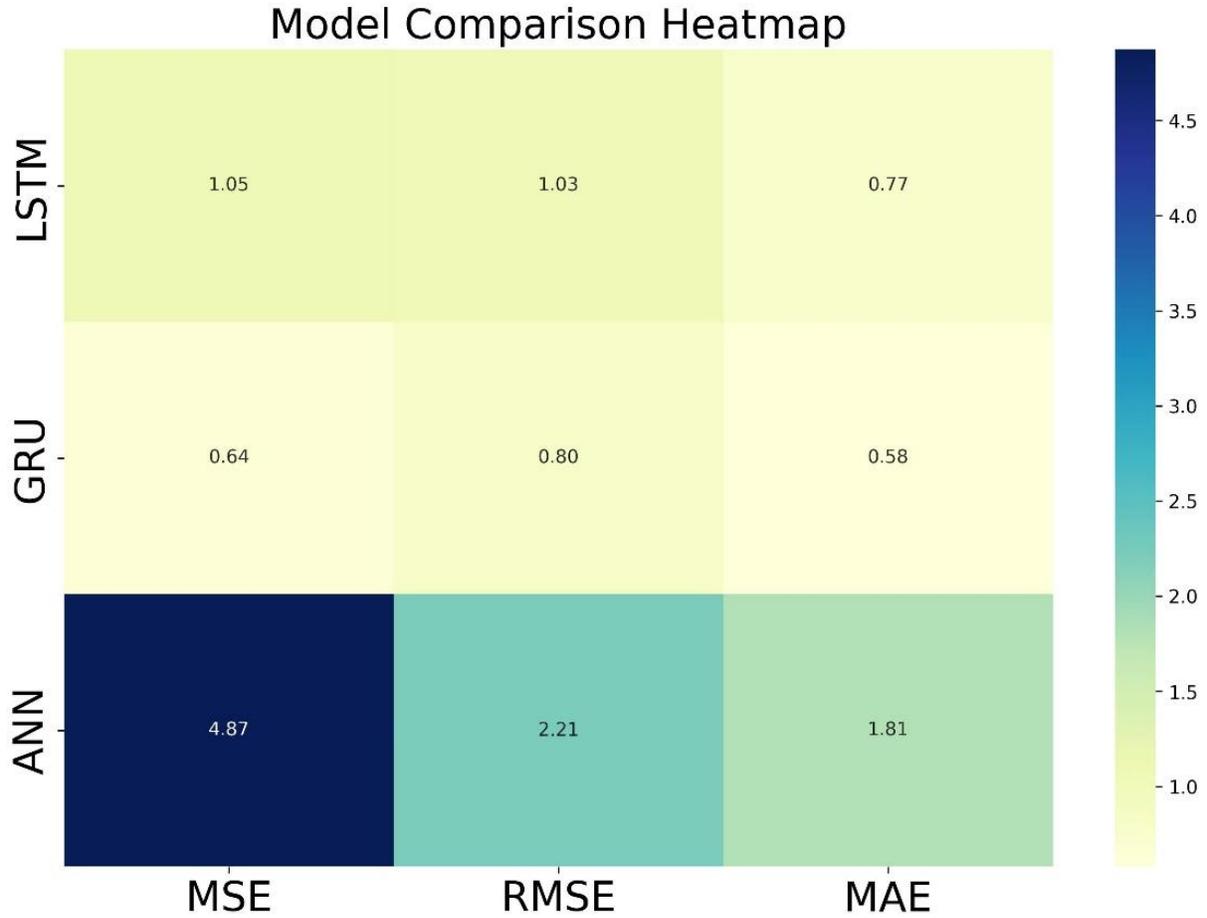

**Figure 11:** Quantitative comparison of deep learning model performance. A heatmap showing MSE, RMSE, and MAE values for LSTM, GRU, and ANN models in predicting battery SOH, providing a comprehensive view of each architecture's effectiveness for this application. The color scale indicates error magnitude, with darker colors representing higher error values.

$$\text{MSE} = \frac{1}{n}\sum_{i=1}^{n}(SOH_{Qi} - SOH'_{Qi})^2 \qquad 9$$

$$RMSE = \sqrt{\frac{1}{n}\sum_{i=1}^{n}(SOH_{Qi} - SOH'_{Qi})^2} \qquad 10$$

$$\text{MAE} = \frac{1}{n}\sum_{i=1}^{n} \mid SOH_{Qi} - SOH'_{Qi} \mid \qquad \qquad 11$$

Where $SOH_{Qi}$ is the actual state of health of the battery, $SOH'_{Qi}$ is the estimated state of health of the battery from the model and n is the total number of data samples.

The relatively low RMSE values for GRU and LSTM (below 1.1) suggest good prediction accuracy, making these recurrent models promising candidates for real-world battery management systems. However, the ANN's higher RMSE (2.21) indicates that it may not be suitable for precise SOH predictions without further refinement. These results provide a clear quantitative basis for model selection. They highlight the advantages of recurrent architectures (GRU and LSTM) over feedforward networks (ANN) for this application. The superior performance of GRU and LSTM aligns with our understanding of battery degradation as a complex, history-dependent process, as revealed in our earlier analyses of parameter evolution over time **(Figures 3-7)**. This comparison guides our choice of model for battery health prediction and reinforces the importance of considering temporal dynamics in battery management strategies. Our analyses reveal that the simpler GRU architecture may offer a better balance of complexity and performance for predicting battery SOH.

# Conclusion

This study presents a comprehensive multi-parameter analysis of Li-ion battery degradation, integrating optical fiber sensing with differential state of health metrics and machine learning techniques. Our approach combines non-invasive sensing, advanced data analysis, and deep learning to provide a nuanced assessment of battery health throughout its lifecycle.

Our key findings include the revelation of complex interactions between electrochemical, mechanical, and thermal processes during battery aging, with mechanical degradation potentially serving as an early indicator of battery aging. Differential analysis of capacity, strain, and temperature with respect to voltage provided unique insights into battery behavior, identifying critical events such as phase transitions and structural changes. Our proposed differential intensity-based State of Health (SOH) metrics showed strong correlations with traditional capacity-based SOH while capturing additional aspects of degradation.

The application of deep learning models demonstrated varying degrees of accuracy in predicting battery SOH, highlighting the importance of model selection and the consideration of temporal dependencies in degradation processes. Notably, our analysis revealed that the Gated Recurrent Unit (GRU) model outperformed both Long Short-Term Memory (LSTM) and Artificial Neural Network (ANN) models across multiple performance metrics. This finding challenges the assumption that more complex recurrent architectures like LSTM always provide superior performance in temporal sequence modeling tasks. The GRU model's superior performance, followed closely by LSTM, underscores the significance of capturing temporal dependencies in battery degradation processes. The substantially higher error rates observed in the ANN model emphasize the limitations of simple feedforward architectures in modeling the complex, time-dependent nature of battery health evolution.

While our approach shows promise, it is important to note its limitations. The study was conducted on a specific type of Li-ion button cell, and the applicability of these findings to other cell chemistries and form factors requires further investigation. Additionally, the long-term stability and reliability of the optical sensing system in real-world applications need to be validated.

Future work should focus on extending this approach to different battery chemistries and form factors and integrating these techniques with emerging battery technologies. Further research is also needed to translate these findings into practical battery management systems that can operate in real time and under varying environmental conditions.

This study demonstrates the power of combining advanced sensing, innovative data analysis, and machine learning in battery diagnostics, contributing significantly to improving the reliability, safety, and sustainability of energy storage systems. The multi-parameter approach presented here can enhance our understanding of battery degradation mechanisms and improve the accuracy of battery health prediction, leading to more efficient and safer energy storage solutions.